# Characterizing two-dimensional incompressible flows through traveling wave and symmetry solutions of the vorticity transportation equation

Lang Xia

Email: langxia.org@gmail.com

**Abstract:** We use the vorticity transportation equation as the start point--with the help of stream function for two-dimensional planar incompressible flows--to obtain exact solutions that characterize evolution and dynamics of the flows. These traveling wave solutions, including both continuous and solitary waves, also recover the results existing in the theory of potential flows. We further analyze the flow patterns regarding symmetry solutions generated by Lie transformation groups. Some of the symmetry solutions have not been discussed in the current literature. They could display flow patterns very different from that of the original exact solutions, suggesting the symmetry solutions are not trivial. Particularly, the symmetry solutions to Lame-Oseen vortex and doublet exhibit more complicated dynamics.

## Introduction

The Navier-Stokes equation (NSE) in classical mechanics is probably the most important PDE. Due to its complexity, neither the uniqueness nor existence of global solutions has been proved mathematically till now [1]. In the engineering fields[2], research in the fluid mechanics whose phenomena are governed by the NSE keeps advancing with the help of numerical computations [3-6]. However, From the theoretical aspect, exact solutions of the NSE are still precious because exact solutions provide a gauge for the verification and validation of all kind of numerical schemes and algorithms [7].

The effort of searching for exact solutions of the NSE has never ceased. By imposing specific boundary and initial conditions, many exact solutions of practical significance originated in the engineering field have been obtained and documented [8-11]. Traditional techniques of solving general PDE has also applied to search exact solutions of NSE [8, 12-14]. Meanwhile, more general solving ansatz such as Lie group method has caught much interest in reducing and solving the NSE and the corresponding variant forms [15-17]. In the present paper, we are trying to obtain some exact solutions using the classical technique. If the solutions are obtained, we then treat it as “seed” solutions. With the assistance of the Lie transformation groups associated with the NSE, we are going to generate symmetry solutions; thereby we are able to study the characteristics of the flow using the new solutions. More specifically, we are going to characterize the two-dimensional flow patterns using exact solutions. Firstly, By the employment of the stream function, we rewrite the two-dimensional NSE in a simple scalar equation. Then we obtain exact solutions to the equation using traveling wave solving method. The traveling wave solutions are finally discussed using the Lie group analysis. We study the dynamics of two-dimensional incompressible flow via both the exact and symmetry solutions to the unsteady equation of stream function.

## Basic equations

We may begin with presenting a logical, rather than rigorous mathematical derivation of the vorticity transportation equation for the sake of easier reading. Let consider the 2-D non-dimensional NSE without external forces in a simply connected domain that is written in the form of

$$\begin{aligned}&\frac{\partial \mathbf{u}}{\partial t}+\mathbf{u}\cdot\nabla\mathbf{u}=-\nabla p+\frac{1}{\mathrm{Re}}\nabla^2\mathbf{u}\\&\nabla\cdot\mathbf{u}=0\end{aligned}\tag{1}$$

where $p$ is the mechanical pressure, Re the Reynolds number and the velocity $\mathbf{u}=(u,v)\in C^\infty(\mathbb{R}^2)$ . Cross product the above equation, we may obtain

$$\frac{\partial \omega}{\partial t}+u\frac{\partial \omega}{\partial x}+v\frac{\partial \omega}{\partial y}=\frac{1}{Re}\nabla^2\omega\tag{2}$$

where

$$\omega=\frac{\partial v}{\partial x}-\frac{\partial u}{\partial y}\tag{3}$$

Because of the incompressible condition, we may introduce the follow stream function $\psi(t,x,y)$ , such that

$$u=\frac{\partial \psi}{\partial y},\quad v=-\frac{\partial \psi}{\partial x}\tag{4}$$

Therefore, the vorticity equation can be written into the following compact form

$$\frac{\partial}{\partial t}\left(\nabla^2\psi\right)+\frac{\partial\psi}{\partial y}\frac{\partial}{\partial x}\left(\nabla^2\psi\right)-\frac{\partial\psi}{\partial x}\frac{\partial}{\partial y}\left(\nabla^2\psi\right)=\frac{1}{Re}\nabla^4\psi \tag{5}$$

where the 2-D biharmonic function is in the form of

$$\nabla^4\psi=\frac{\partial^4\psi}{\partial x^4}+\frac{\partial^4\psi}{\partial y^4}+2\frac{\partial^4\psi}{\partial x^2\partial y^2} \tag{6}$$

The 4$^{th}$ order scalar equation Eq.(5) offers us simplicity to carry out symmetry or numerical analysis. It gives us a simpler expression for the dynamics of the vector potential field of fluids. The benefit of using such stream function is that the velocity components can be calculated later on by the derivation of the stream function. Also, the pressure will be recovered by

$$p=\int\frac{\partial p}{\partial x}dx+\int\frac{\partial p}{\partial y}dy+p_0 \tag{7}$$

where the partial derivatives of the pressure are determined by Eq.(1).

In engineering application, the solutions of differential equations are not of much interest without specifying the initial and boundary conditions. Four boundary conditions are required to solve Eq.(5). For the following general analysis, the initial and boundary condition will be imposed naturally into the analysis and solving process.

## Solution Ansatz

### *A. Traveling wave solutions*

Since the nonlinear equation derived in the previous section is nothing but a type of evolution equation containing the diffusion and convection, we thus try to pursue traveling wave solutions to the equations by assuming

$$\psi(x,y,t)=\varphi(z) \tag{8}$$

with $z=k_1x+k_2y+d_0-\varpi t$ . Here we may call $k_1$, $k_2$ wave numbers, $\varpi$ the angular frequency. $d_0$ the phase constant. Note that all the above coefficients can be complex numbers. If the equation admits the traveling waves solution, the wave speed can be determined by $c=\varpi/k$ . Substitute the Eq.(8) into Eq.(5), we obtain the following linear ordinary differential equation

$$\left(k_2^2+k_1^2\right)\left(\frac{d^4}{dz^4}\phi(z)\right)-\mathrm{Re}\,\varpi\left(\frac{d^3}{dz^3}\phi(z)\right)=0 \tag{9}$$

If $k_1^2+k_2^2\neq 0$, a solution is readily obtained in the form of

$$\varphi(\mathrm{z})=C_1\exp(-\frac{\varpi\mathrm{Re}}{k_2^2+k_1^2}z)+C_2z^2+C_3z+C_4 \tag{10}$$

where the integration constants $C_i$ is determined by boundary conditions.

When $k_2^2+k_1^2=0$ , we have a solution for an arbitrary function satisfying

$$\psi(x,y,t)=\varphi(kx+jky+d_0-\varpi t) \tag{11}$$

where $j=\sqrt{-1}$.

Actually, in this case, the arbitrary function satisfies the following Laplace equation

$$\nabla^2\varphi=0 \tag{12}$$

which is nothing but the solution of planar irrotational flows with the vorticity $\omega=0$. It is usually studied in the framework of potential flow theory, in which the Reynolds number does not come into the solutions [18]. Thus, considering the kinematics of the flow is enough to characterize the very nature. In this situation, the traveling wave solutions are identical to the potential flow theory, providing that $k_1=k_2=1,\ d_0=\varpi=0$ and the imaginary part is the potential here. The linearity of the Laplace equation also suggests that the superposition of its solutions is still the solution of the Eq.(5). Therefore, we can easily extend the analysis on the potential flow of the traveling wave solutions. In other words, the traveling wave solutions generalize the classical potential flow theory with the possibility of altering the evolution of the potential flow. Following the logical of potential flow, we may continue the analysis with analogy to the irrotational flow of some elementary cases e.g. $F=A\cdot z^n$ , that is

$$\psi(x,y,t)=A\cdot(kx+jky+d_0-\varpi t)^n \tag{13}$$

where $A$ is an arbitrary constant. The real part of Eq.(13) is the stream function. Particularly, the interesting cases may be discussed by choosing different values of $n$. Remember also the linearity of Eq.(12) suggesting that we can construct more complex flows using the traveling wave solutions, but it is not the scope of the present paper.

### *B. Symmetry Solutions*

Enlightened by the success of Galois groups in solving algebraic equations, Sophus Lie proposed continuous groups for solving differential equations [19, 20]. This method helped Lie to discover the symmetries hidden in the differential equations. Meanwhile, it unified many kinds of *ad hoc* methods of reducing differential equations. Although Lie groups have richer meanings in the fields of differential geometry and mathematical physics, It is only after 1970 they become a power tool to simplify and solve differential equations [21, 22]. That is because the Lie group method requires an enormous amount of algebraic calculations, and sometimes solving the linear determining equations are even harder than solving the original equations. Thanks to the developments of modern computer systems, those determining equations can be easily solved with the help of CAS software or packages right now. Therefore, within the past 30 years, the symmetric groups, Lie algebras and the corresponding invariant solutions to many famous equations have been gradually documented [2, 17, 23].

In order to generate new exact solutions, we need to employ the technique above of Lie group analysis, of which reduction or solution of the Eq.(5) could be achieved. Recall the one-parameter Lie transformation group of the form [24]

$$\begin{aligned} x^* &= x+\varepsilon\xi_1(x,y,t,\psi)+O(\varepsilon^2) \\ y^* &= y+\varepsilon\xi_2(x,y,t,\psi)+O(\varepsilon^2) \\ t^* &= t+\varepsilon\xi_3(x,y,t,\psi)+O(\varepsilon^2) \\ \psi^* &= \psi+\varepsilon\eta_1(x,y,t,\psi)+O(\varepsilon^2) \end{aligned} \tag{14}$$

and the corresponding infinitesimal generator

$$X=\xi_1\frac{\partial}{\partial x}+\xi_2\frac{\partial}{\partial y}+\xi_3\frac{\partial}{\partial z}+\xi_4\frac{\partial}{\partial t}+\eta\frac{\partial}{\partial \phi} \tag{15}$$

where

$$\begin{aligned} \xi_i &= \xi_i\left(x^i,\psi\right)=\frac{dx^{i*}}{d\varepsilon}|_{\varepsilon=0} \\ \eta &= \eta\left(x^i,\psi\right)=\frac{d\psi^*}{d\varepsilon}|_{\varepsilon=0} \end{aligned} \tag{16}$$

and denote $(x^i)=(x,y,z,t),\ i=1,2,3,4$ . Also, the second prolongation (lifting) of the vector field $X$ can be written as

$$X^{(2)}=X+\eta^i\frac{\partial}{\partial\psi_i}+\eta^{ji}\frac{\partial}{\partial\psi_{ji}} \tag{17}$$
.

Here $\psi_{ij}$ denotes the partial derivatives with respective to the indices. By applying

$$X^{(2)}F\left(x^i,\psi_j,\psi_{ji}\right)|_{F=0}=0 \tag{18}$$

the determining equations for $\xi_i$ and $\eta$ can be obtained. By using CAS software or related packages, we can readily solve the systems of linear partial differential equations. For the scalar equation Eq.(5), the determining equations are obtained and solved by using MAPLE. Finally, we can get the infinitesimals are in the following forms:

$$\begin{aligned}
\xi_1 &= \frac{1}{2}C_1x + C_3yt + C_4y + \alpha(t) \\
\xi_2 &= \frac{1}{2}C_1y - C_3xt - C_4x + \beta(t) \\
\xi_3 &= C_1t + C_2 \\
\eta &= \frac{1}{2}C_3\left(x^2+y^2\right) + y\frac{d}{dt}\alpha(t) - x\frac{d}{dt}\beta(t) + \gamma(t)
\end{aligned} \tag{19}$$

where $C_i$ are constant, and the Greek alphabets are arbitrary functions. By setting each constant coefficient to be one and all others to be zero sequentially, the Lie subalgebras of the equation are obtained and spanned by

$$\begin{aligned}
X_1 &= \frac{1}{2}x\frac{\partial}{\partial x} + \frac{1}{2}y\frac{\partial}{\partial y} + t\frac{\partial}{\partial t},\ X_2 = \frac{\partial}{\partial t} \\
X_3 &= yt\frac{\partial}{\partial x} - xt\frac{\partial}{\partial y} + \frac{1}{2}\left(x^2+y^2\right)\frac{\partial}{\partial \psi} \\
X_4 &= y\frac{\partial}{\partial x} - x\frac{\partial}{\partial y}
\end{aligned} \tag{20}$$

As well as the following infinite subalgebras

$$\begin{aligned}
X_\alpha &= \alpha\frac{\partial}{\partial x} + y\dot{\alpha}\frac{\partial}{\partial \psi}, \\
X_\beta &= \beta\frac{\partial}{\partial y} - x\dot{\beta}\frac{\partial}{\partial \psi}, \\
X_\gamma &= \gamma\frac{\partial}{\partial \psi},
\end{aligned} \tag{21}$$

.

Therefore, the transformation groups can be calculated by using Eq.(16)

$$
\begin{aligned}
&G_1 : (x, y, t, \psi) \mapsto (xe^{\frac{\varepsilon}{2}}, ye^{\frac{\varepsilon}{2}}, te^{\varepsilon}, \psi) \\
&G_2 : (x, y, t, \psi) \mapsto (x, y, t+\varepsilon, \psi) \\
&G_3 : (x, y, t, \psi) \mapsto (x+\varepsilon yt, y-\varepsilon xt, t, \psi+\varepsilon \frac{x^2+y^2}{2}) \\
&G_4 : (x, y, t, \psi) \mapsto (x+\varepsilon y, y-\varepsilon x, t, \psi)
\end{aligned} \tag{22}
$$

as well as

$$
\begin{aligned}
&G_\alpha : (x, y, t, \psi) \mapsto (x+\varepsilon\alpha(t), y, t, \psi+\varepsilon\dot{\alpha}(t) y) \\
&G_\beta : (x, y, t, \psi) \mapsto (x, y+\varepsilon\beta(t), t, \psi-\varepsilon\dot{\beta}(t) x) \\
&G_\gamma : (x, y, t, \psi) \mapsto (x, y, t, \psi+\varepsilon\gamma(t))
\end{aligned} \tag{23}
$$

The above transformation groups enable us to generate symmetry solutions from the traveling wave solutions. If $\psi(x, y, t)$ is the solution of the Eq.(5), so as the following solutions

$$
\begin{aligned}
&\psi(xe^{-\frac{\varepsilon}{2}}, ye^{-\frac{\varepsilon}{2}}, te^{-\varepsilon}), \;\; \psi(x, y, t-\varepsilon) \\
&\psi(x-\varepsilon yt, y+\varepsilon xt, t)+\varepsilon \frac{x^2+y^2}{2}, \;\; \psi(x-\varepsilon y, y+\varepsilon x, t)
\end{aligned} \tag{24}
$$

and

$$
\begin{aligned}
&\psi(x-\varepsilon\alpha(t), y, t)+\varepsilon\dot{\alpha}(t) y \\
&\psi(x, y-\varepsilon\beta(t), t)-\varepsilon\dot{\beta}(t) x \\
&\psi(x, y, t)+\varepsilon\gamma(t)
\end{aligned} \tag{25}
$$

In the present work, we only focus on the $G_3$ of Eq.(22), which is an unsteady case of $G_4$, but with stretching. $G_1$, $G_2$, $G_4$ are trivial symmetries in this point of view because the flow pattern may not depend on the coordinate and thus scaling, translations and rotations could not alter the dynamical flavors of the flow. As for the transformation groups generated by the infinite sub-algebras, they also belong to the translations. Therefore, we may have the following symmetry solutions by using $G_4$

$$
\psi(x, y, t) = \varphi\left(k(x-\varepsilon yt)+jk(y+\varepsilon xt)+d_0-\varpi t\right)+\varepsilon \frac{x^2+y^2}{2} \tag{26}
$$

$$\psi(x,y,t) = C_1 \exp(-\frac{\varpi \mathrm{Re}}{k_2^2 + k_1^2}(k_1(x-\varepsilon yt) + k_2(y+\varepsilon xt) + d_0 - \varpi t))$$
$$+C_2(k_1(x-\varepsilon yt) + k_2(y+\varepsilon xt) + d_0 - \varpi t))^2 \tag{27}$$
$$+C_3(k_1(x-\varepsilon yt) + k_2(y+\varepsilon xt) + d_0 - \varpi t)) + C_4 + \varepsilon \frac{x^2+y^2}{2}$$

In addition to the above symmetry solution, the Lie Group method also generated the following invariant solutions by the virtue of corresponding Lie subalgebras

$$\psi\left(x,y,t\right) = C_1 \int^{\frac{t}{y^2+x^2}} \frac{\ln(z) - \mathrm{Ei}\left(1, -\frac{\mathrm{Re}}{4z}\right) e^{-\frac{\mathrm{Re}}{4z}}}{z} dz + C_2 \ln\left(\frac{t}{y^2+x^2}\right) + \mathrm{Ei}\left(1, \frac{\mathrm{Re}\left(y^2+x^2\right)}{4t}\right) C_3 + C_4$$

(28)

where the velocity vector is given by

$$u = \frac{2C_1 y\left(\mathrm{Ei}\left(1, -\frac{\mathrm{Re}\left(y^2+x^2\right)}{4t}\right) e^{-\frac{\mathrm{Re}\left(y^2+x^2\right)}{4t}} - \ln\left(\frac{t}{y^2+x^2}\right)\right)}{y^2+x^2} - \frac{2C_2 y}{y^2+x^2} - \frac{2C_3 y}{y^2+x^2} e^{-\frac{\mathrm{Re}\left(y^2+x^2\right)}{4t}}$$

$$v = -\frac{2C_1 x\left(\mathrm{Ei}\left(1, -\frac{\mathrm{Re}\left(y^2+x^2\right)}{4t}\right) e^{-\frac{\mathrm{Re}\left(y^2+x^2\right)}{4t}} - \ln\left(\frac{t}{y^2+x^2}\right)\right)}{y^2+x^2} + \frac{2C_2 x}{y^2+x^2} + \frac{2C_3 x}{y^2+x^2} e^{-\frac{\mathrm{Re}\left(y^2+x^2\right)}{4t}}$$

(29)

It is not difficult to see that the second term on the right-hand side of the above equations is an expression of a Rankine Vortex, and the third term is of a Lame-Oseen Vortex (See figure 1. We only plot the Lame-Oseen vortex.)

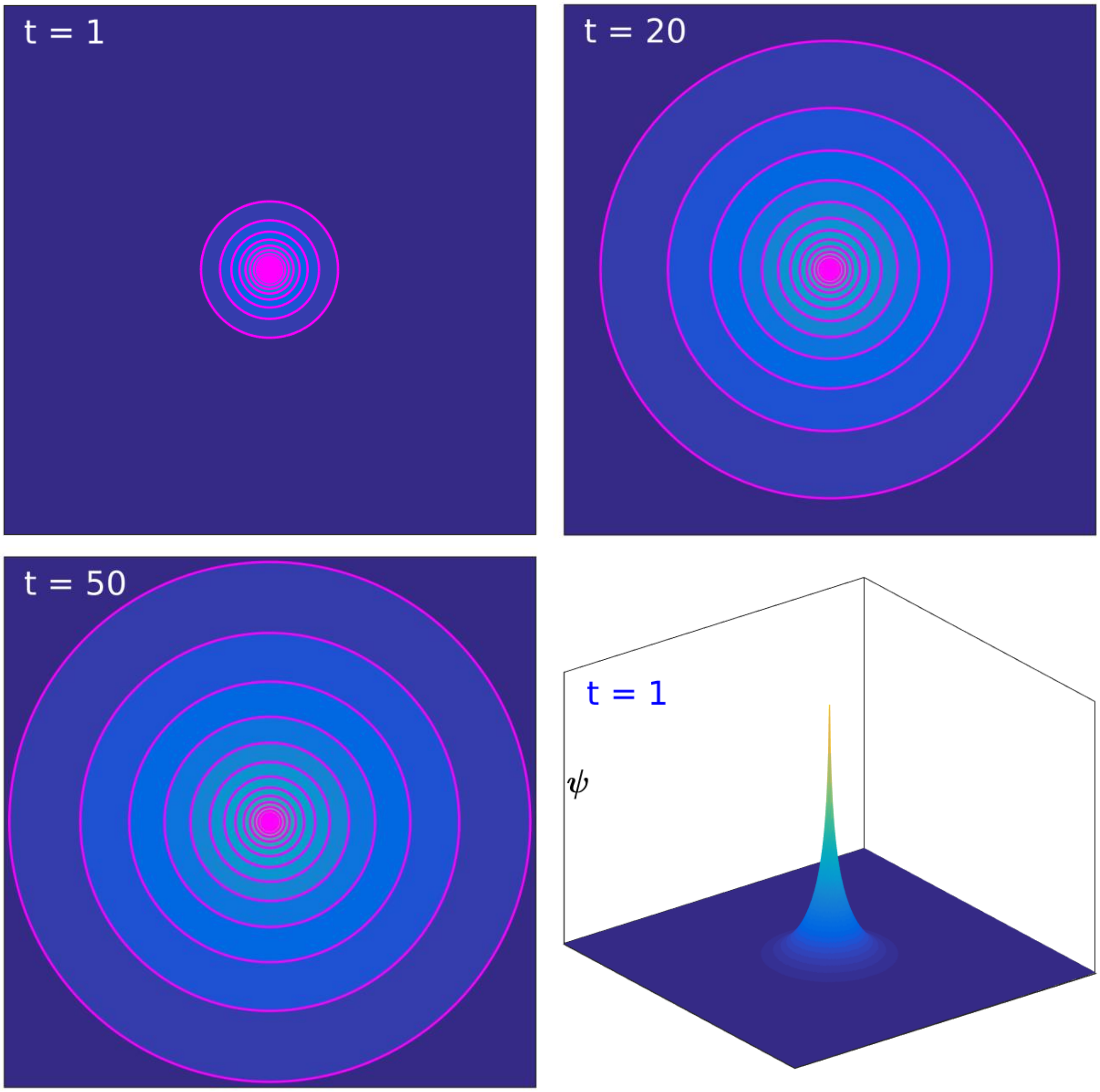

**Figure 1**: Lame-Oseen Vortex.

## Results

For the sake of simplicity, we assume the wave number $k_1 = k_2 = 2$ , the phase constant $d_0 = 0$ and the angular frequency $\varpi = 0.1$ to visualize the flow patterns. They correspond to certain initial and boundary conditions.

### *A. Continuous traveling wave solution*

Using the general solution obtained in the last section, we are able to construct specific solutions for the real case. For a simple case of traveling continuous waves, we may assume

$$\psi(x,y,t) = \cos\left(kx + jky + d_0 - \varpi t\right) \tag{30}$$

Using the following identities

$$\begin{aligned} \cos(jy) &= \frac{e^{-y}+e^{y}}{2} = \cosh(y) \\ \sin(jy) &= \frac{e^{-y}-e^{y}}{2j} = j\sinh(y) \end{aligned} \tag{31}$$

we have

$$\psi(x,y,t) = \cosh\left(ky\right)\cos\left(kx - \varpi t + d_0\right) + j\sinh\left(ky\right)\sin\left(kx - \varpi t + d_0\right) \tag{32}$$

Due to the symmetry of the solution form, we can also obtain

$$\psi(x,y,t) = \cosh\left(kx\right)\cos\left(ky - \varpi t + d_0\right) + j\sinh\left(kx\right)\sin\left(ky - \varpi t + d_0\right) \tag{33}$$

The solution characterizes vortices propagate in $x$ direction with a speed of $c = \varpi / k$ . Note that the amplitude varies only in $y$ direction. Their evolutions are plotted in figure 2. The linear combination of the above solutions will also be the solution of Eq.(12). It is nothing but with vortices being added in $y$ direction. Furthermore, other trigonometric functions can also be applied to generate new solutions.

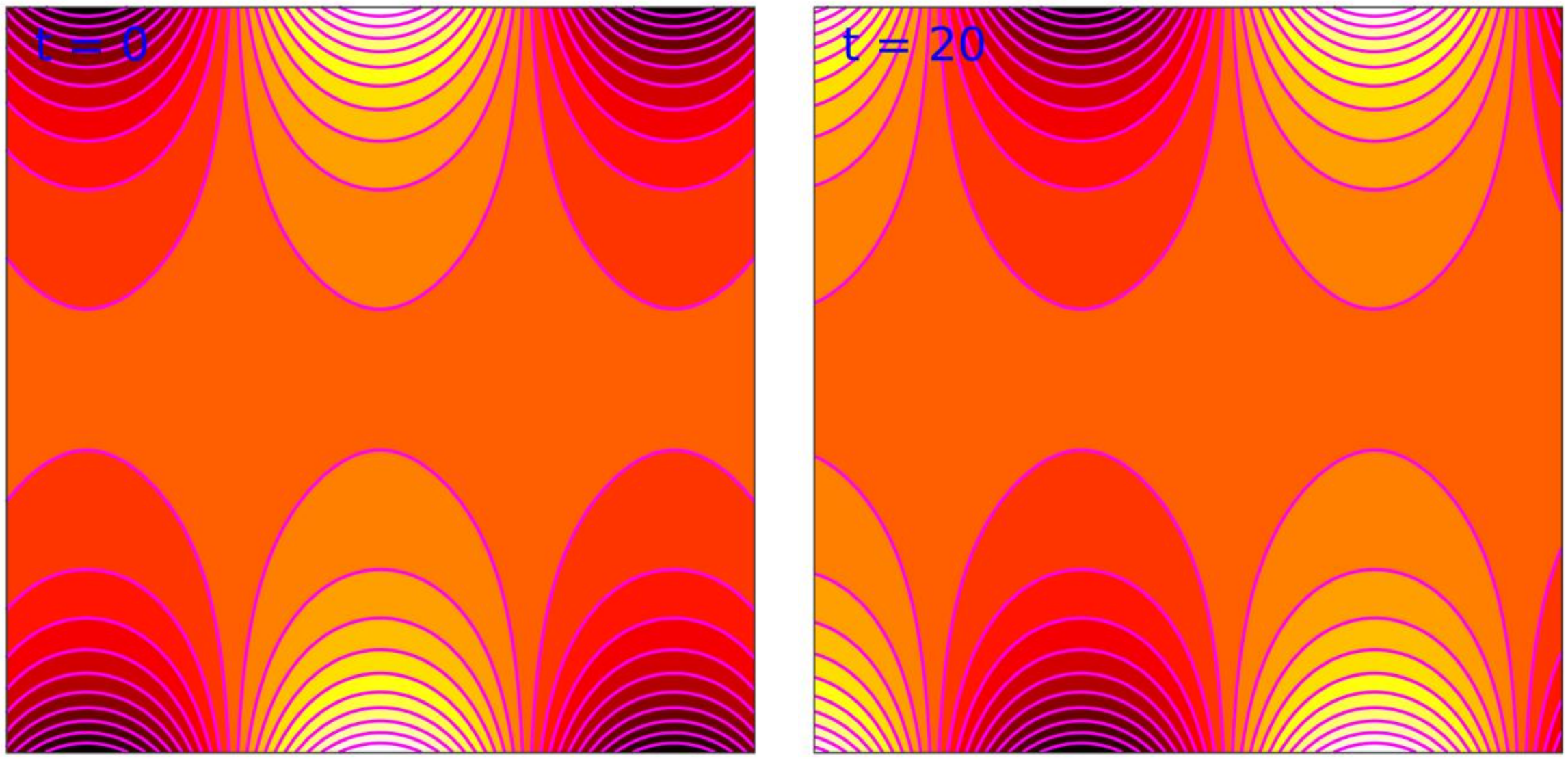

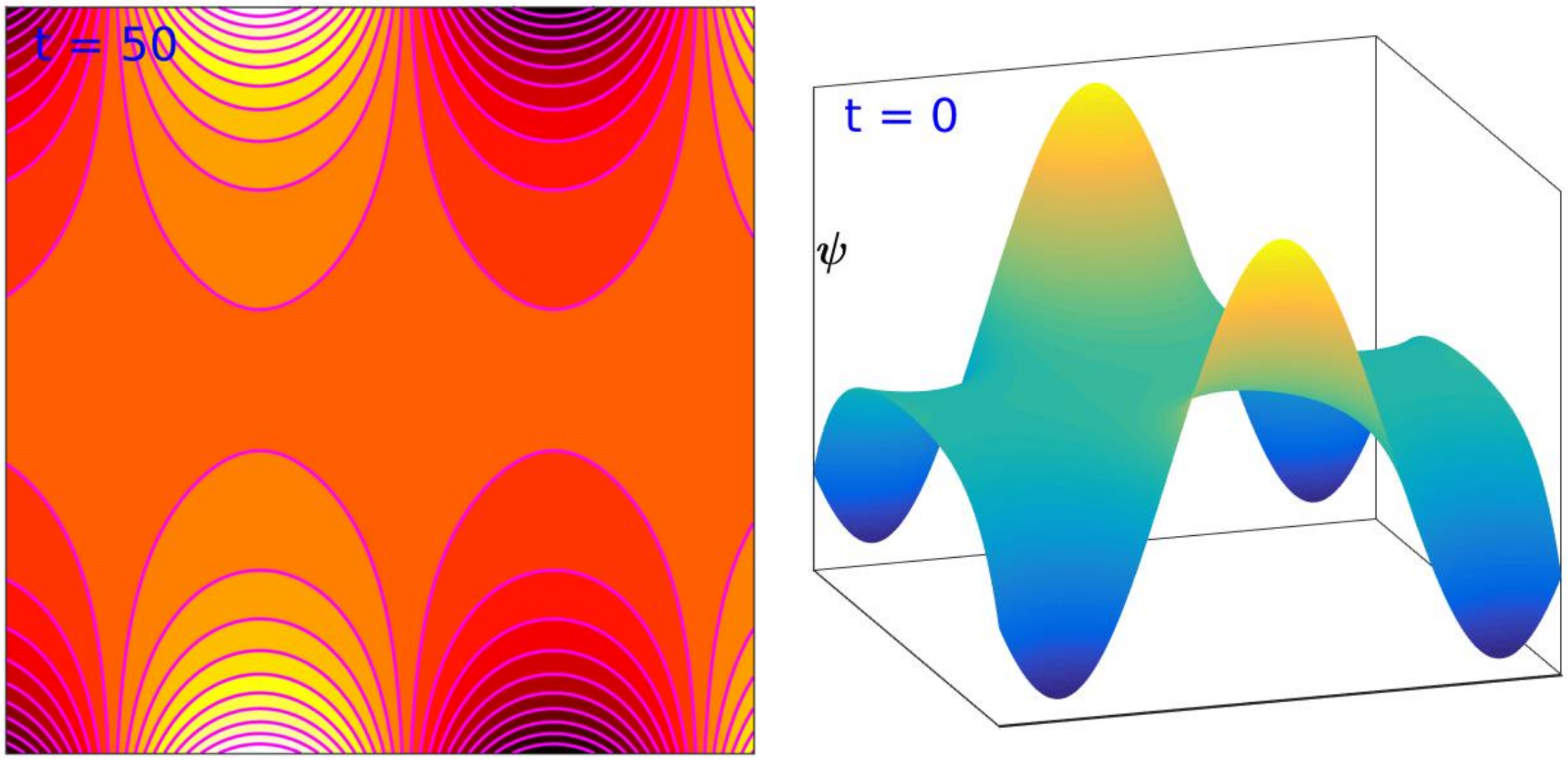

**Figure 2**: Contours of the stream function Eq.(30) at different time and the three-dimensional profile at t = 0.

Applying the Lie group transformation $G_3$ to the real part of Eq.(30) , the corresponding symmetry solution is in the form of

$$\psi(x,y,t)=\cosh\left(ky+\varepsilon kxt\right)\cos\left(kx-\varepsilon kyt-\varpi t+d_0\right)+\varepsilon\frac{x^2+y^2}{2} \tag{34}$$

which again satisfies Eq.(12). Its streamline contours are shown in figure 3.

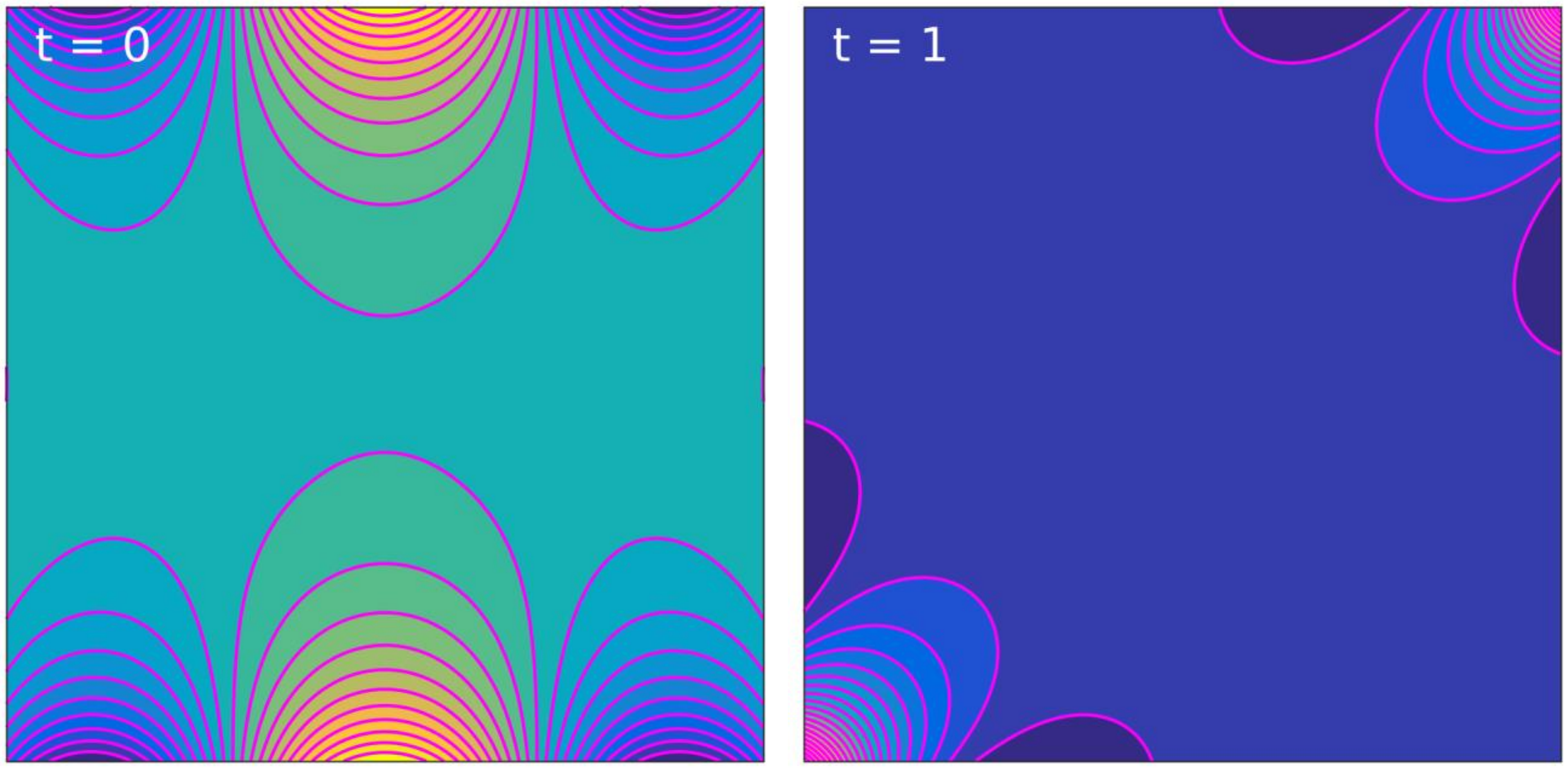

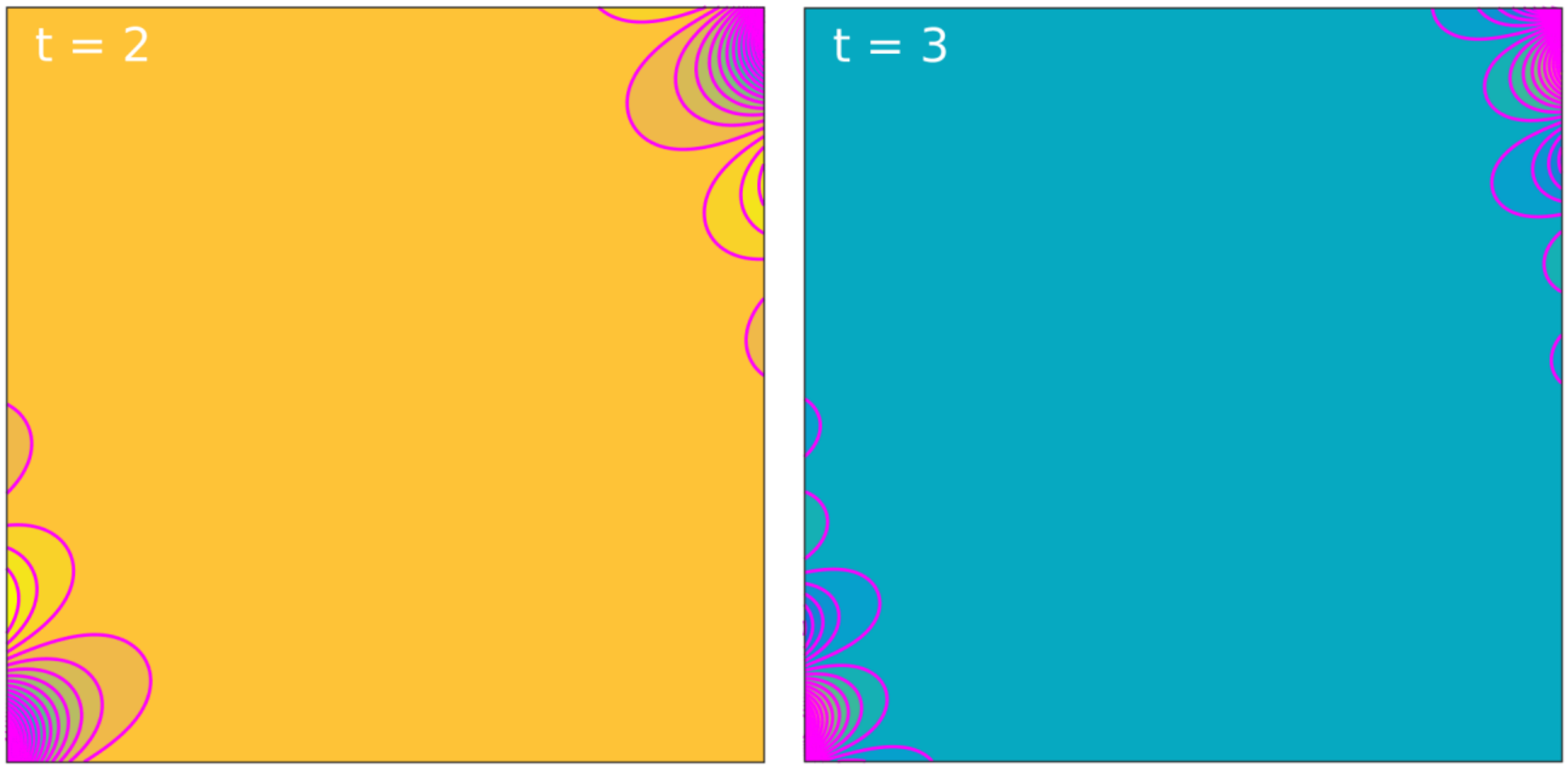

**Figure 3**: Contours of the stream function Eq.(34) at different time.

### *B. Moving Rankine Vortex*

We may also choose a logarithm function, such that

$$\psi(x,y,t)=\ln\left(kx+jky+d_0-\varpi t\right) \tag{35}$$

It is usually called the fundamental solution of the Laplace equation. We can see a moving Rankine vortex, or free vortex (See figure 4). The symmetry solution transformed from the above solution can be written in the form

$$\psi(x,y,t)=\ln\left(k(y+\varepsilon xt)+jk(x-\varepsilon xt)+d_0-\varpi t\right)+\varepsilon\frac{x^2+y^2}{2} \tag{36}$$

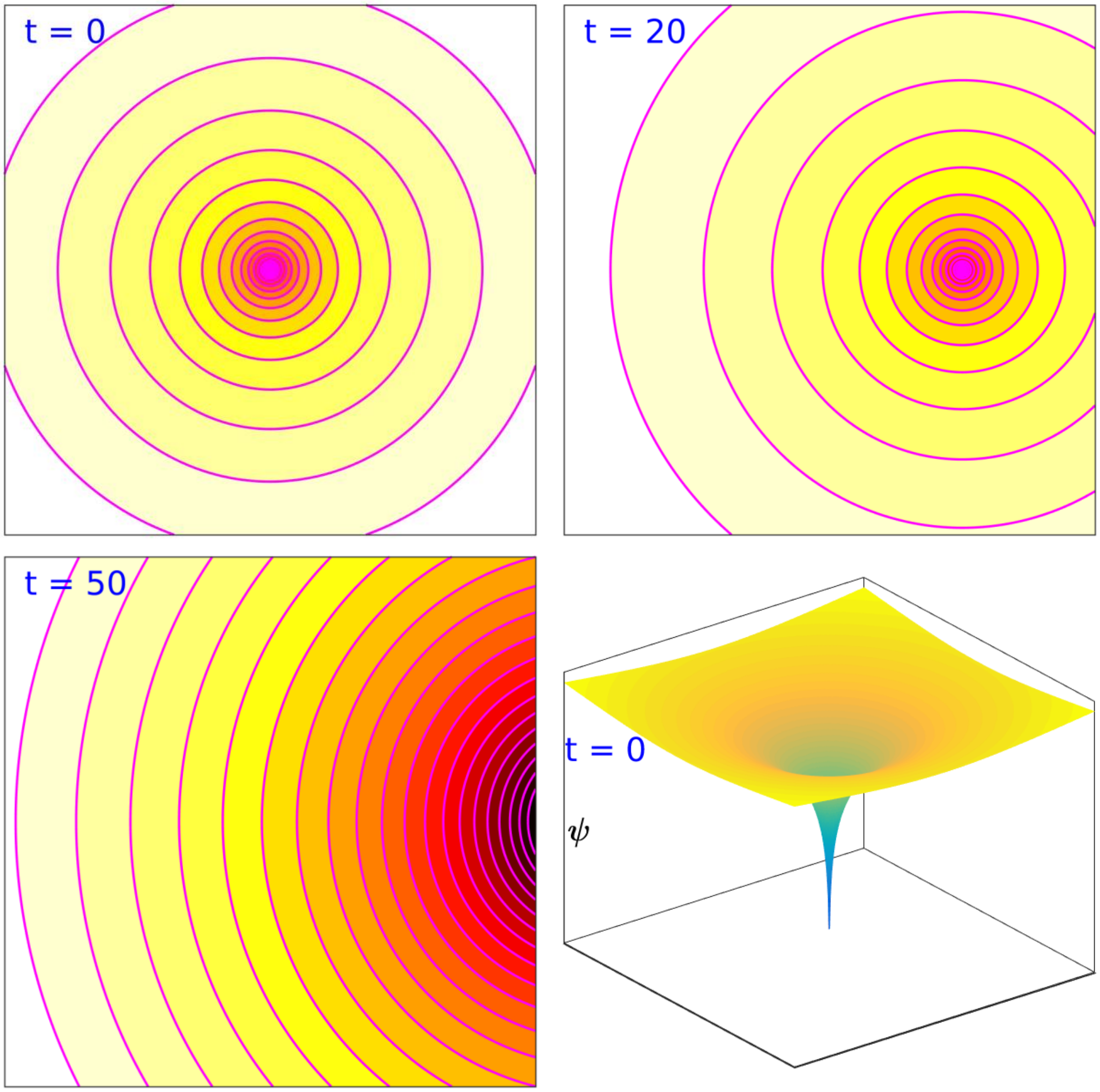

**Figure 4**: Contours of the stream function Eq.(35) at different time and the three-dimensional profile at t = 0.

### *C. Traveling solitary wave solution*

For those partial differential equations that diffusion and convection terms play a major role, the hyperbolic tangent method (tanh method) is a powerful technique to generate exact solutions to those equations [25]. In the present paper, Eq.(5) is still a type of convection-diffusion equation. Thus we may assume it has the traveling solitary wave solution giving rise to the following form

$$\psi(x, y, t) = \tanh\left(kx + jky + d_0 - \varpi t\right) \tag{37}$$

Its flow patterns are plotted in figure 5, which demonstrate two doublets propagate from left to right with a speed of $c = \varpi / k$. Also, recall that a doublet in the potential flow is $\psi(x, y, t) = 1/z$, which depicts a single doublet propagating in $x$ direction.

Using the aforementioned symmetry transformation group, we can get the corresponding symmetry solution in the form of

$$\psi(x, y, t) = \tanh\left(k(y + \varepsilon xt) + jk(x - \varepsilon xt) + d_0 - \varpi t\right) + \varepsilon \frac{x^2 + y^2}{2} \tag{38}$$

The doublets are rotating and stretching as shown in figure 6. It has more complex flow patterns due to the stretching and rotation. In other words, the symmetry solution is able to represent nontrivial solutions of flows.

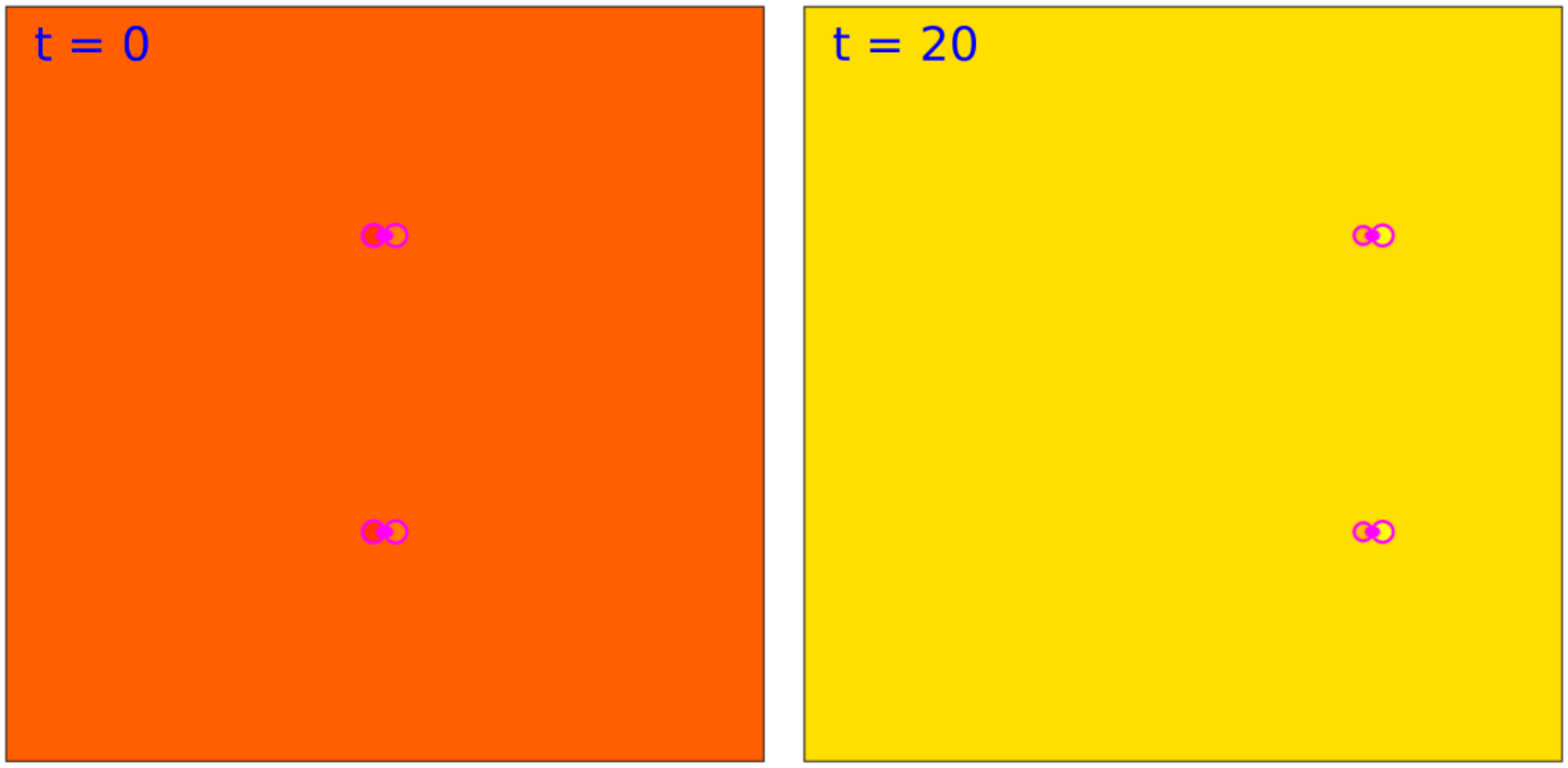

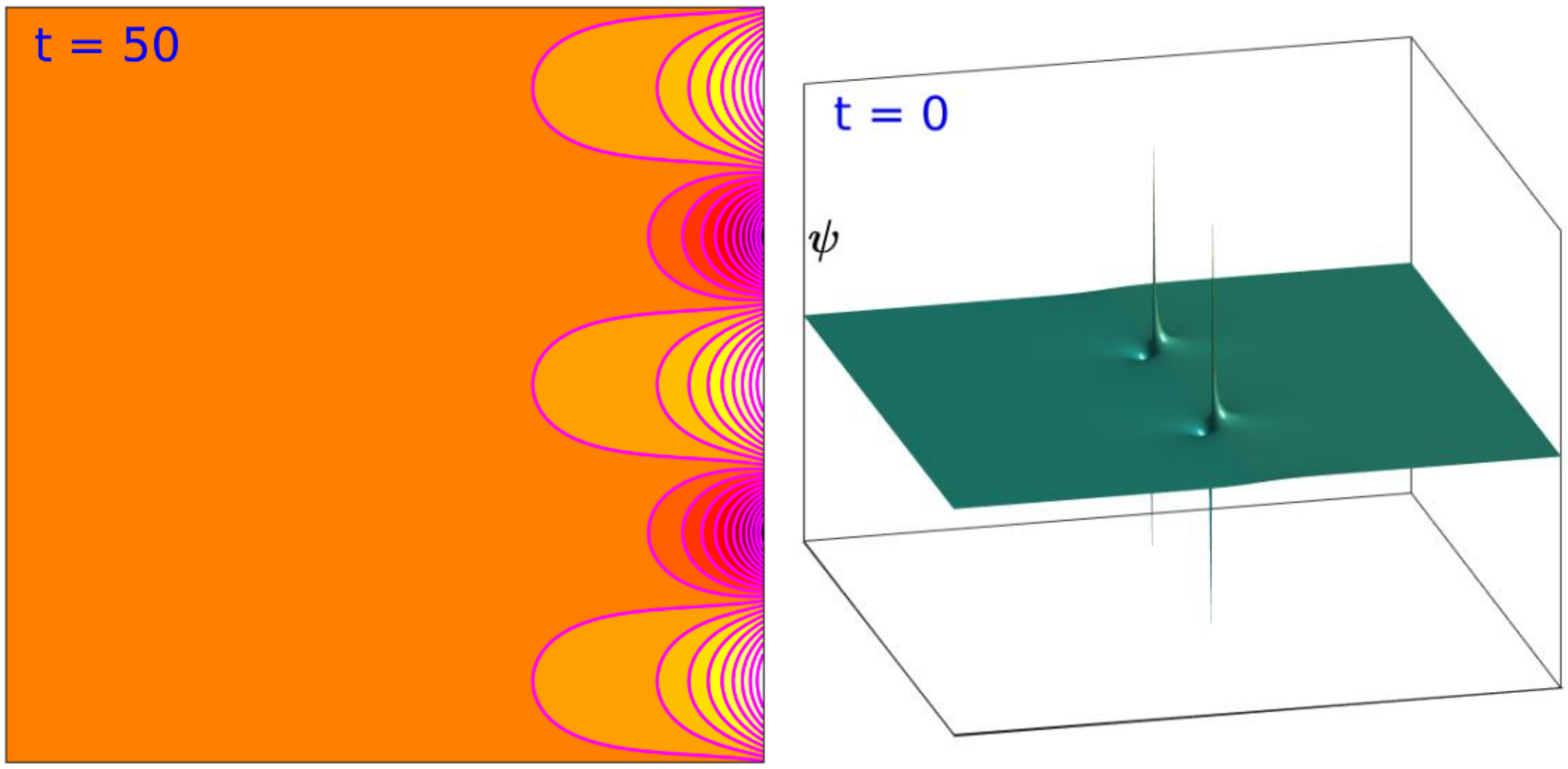

**Figure 5**: Contours of the stream function Eq.(37) at different time and the three-dimensional profile at t = 0.

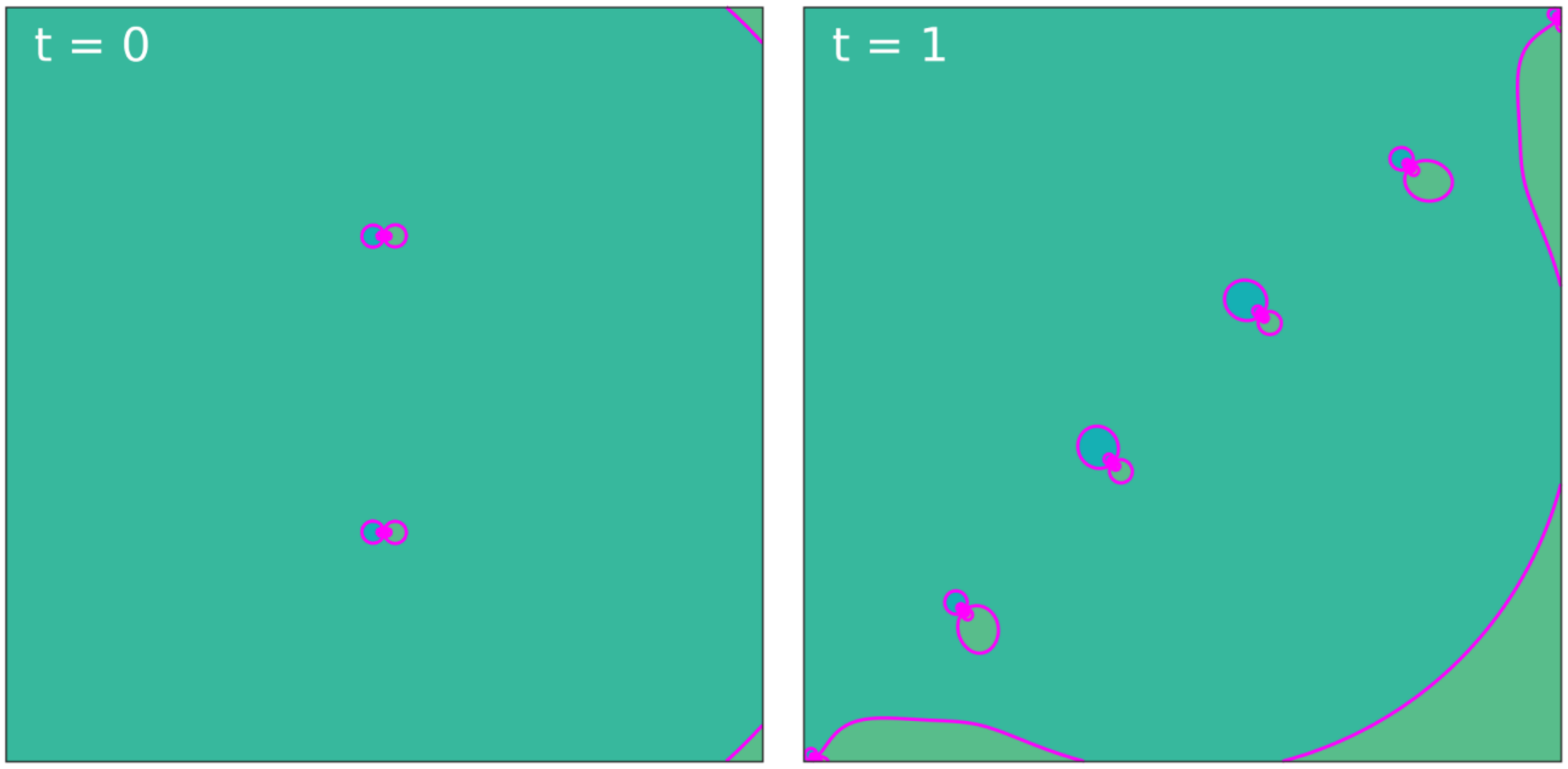

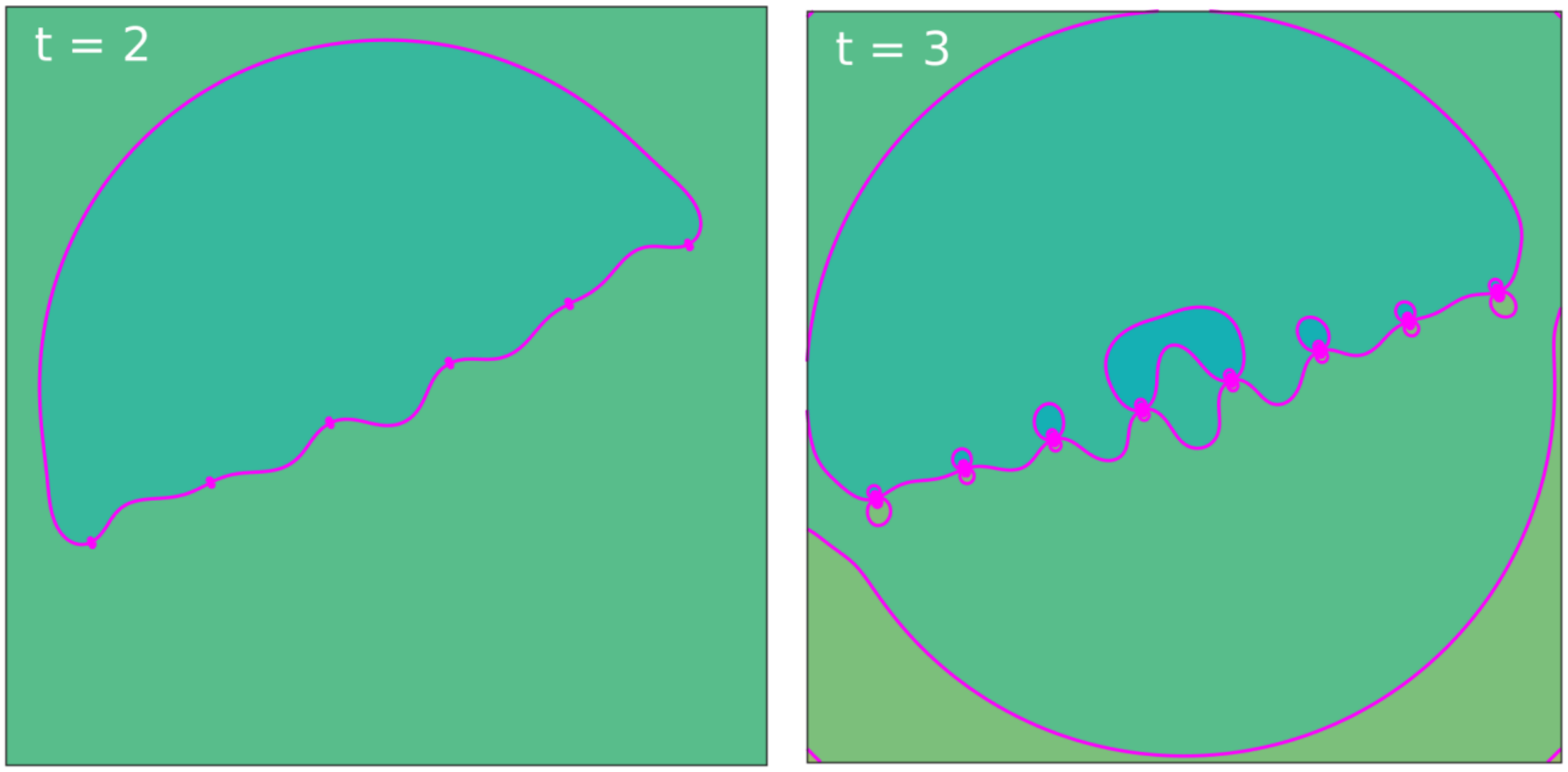

**Figure 6**: Contours of the stream function Eq.(38) at different time.

Note the fact that the solutions of the Laplace equation that are periodic in one direction must be exponential in another direction. We may construct the solution in the form of

$$\begin{aligned} &\psi(x,y,t)=\exp\left(j\left(kx-\varpi t+d_0\right)-ky\right) \\ &or \\ &\psi(x,y,t)=\exp\left(j\left(ky-\varpi t+d_0\right)-kx\right) \end{aligned} \tag{39}$$

All the above solutions are the special case of Eq.(8) and exhibit different vorticity patterns. Therefore, even within the same traveling wave solutions, we may still expect to see the very different dynamic behaviors.

### *D. Solutions of real coefficients*

Besides, we may also obtain the following solutions

$$\begin{aligned} &\psi(x,y,t)=(k_1x+k_2y+d_0-\varpi t)^2 \\ &\psi(x,y,t)=\exp\left(-\frac{\varpi \mathrm{Re}}{k_2^2+k_1^2}(k_1x+k_2y+d_0-\varpi t)\right) \end{aligned} \tag{40}$$

Their patterns are plotted in figure 7 and 9. The contour patterns are very similar to each other, except that the latter behaves like a steady flow. Using the Lie transformation group, we can obtain the corresponding symmetry solutions which are in the form of Eq.(26) and Eq.(27), respectively. Their flow patterns, exhibiting rotations and stretching, are demonstrated in figure 8 and figure 10. More interestingly, a temporary vortex is formed initially in both symmetry solutions. But eventually, the flows go back to their original linearity (linear streamlines).

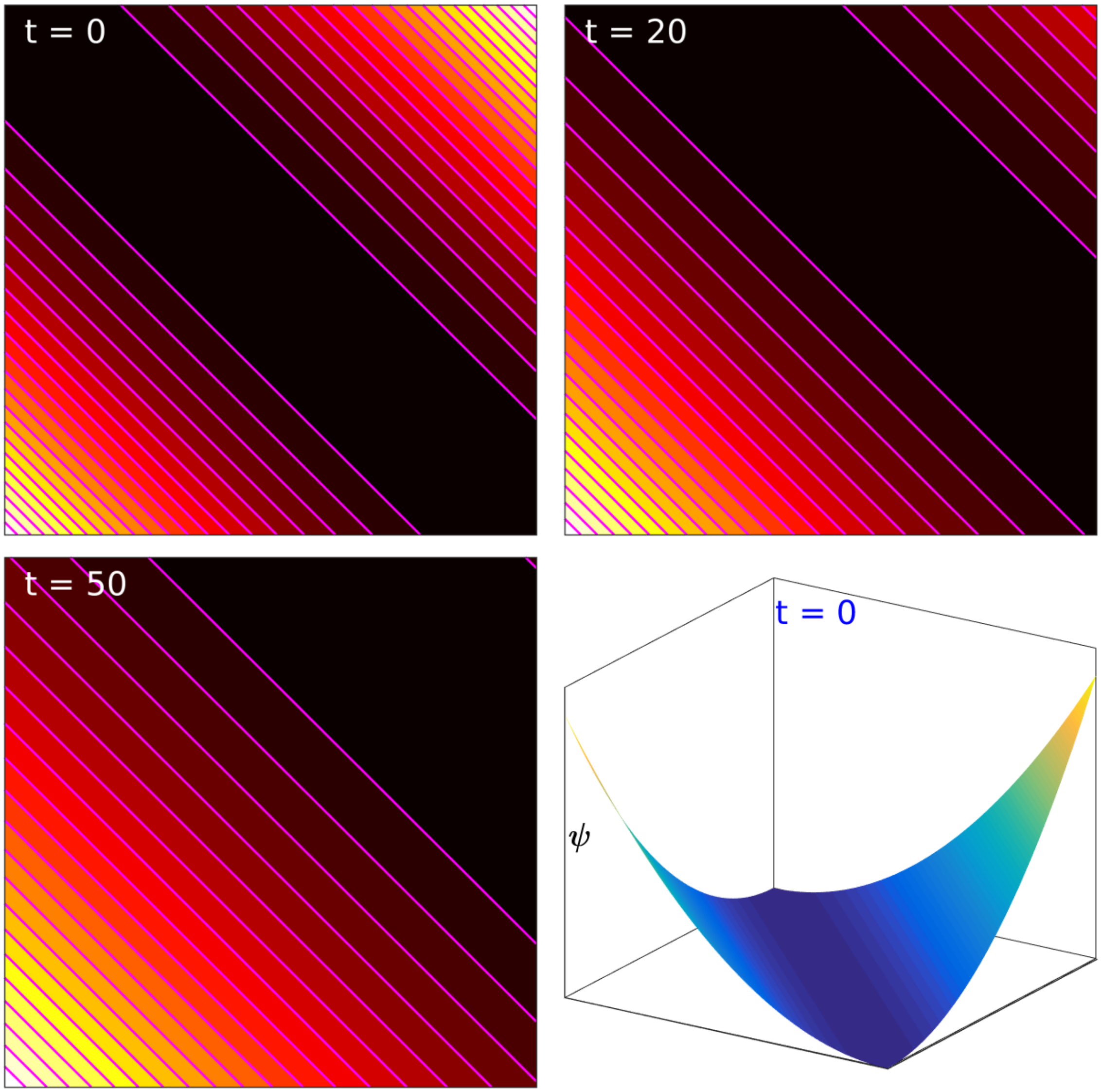

**Figure 7**: Contours of the first stream function in Eq.(40) at different time and the three-dimensional profile at t = 0.

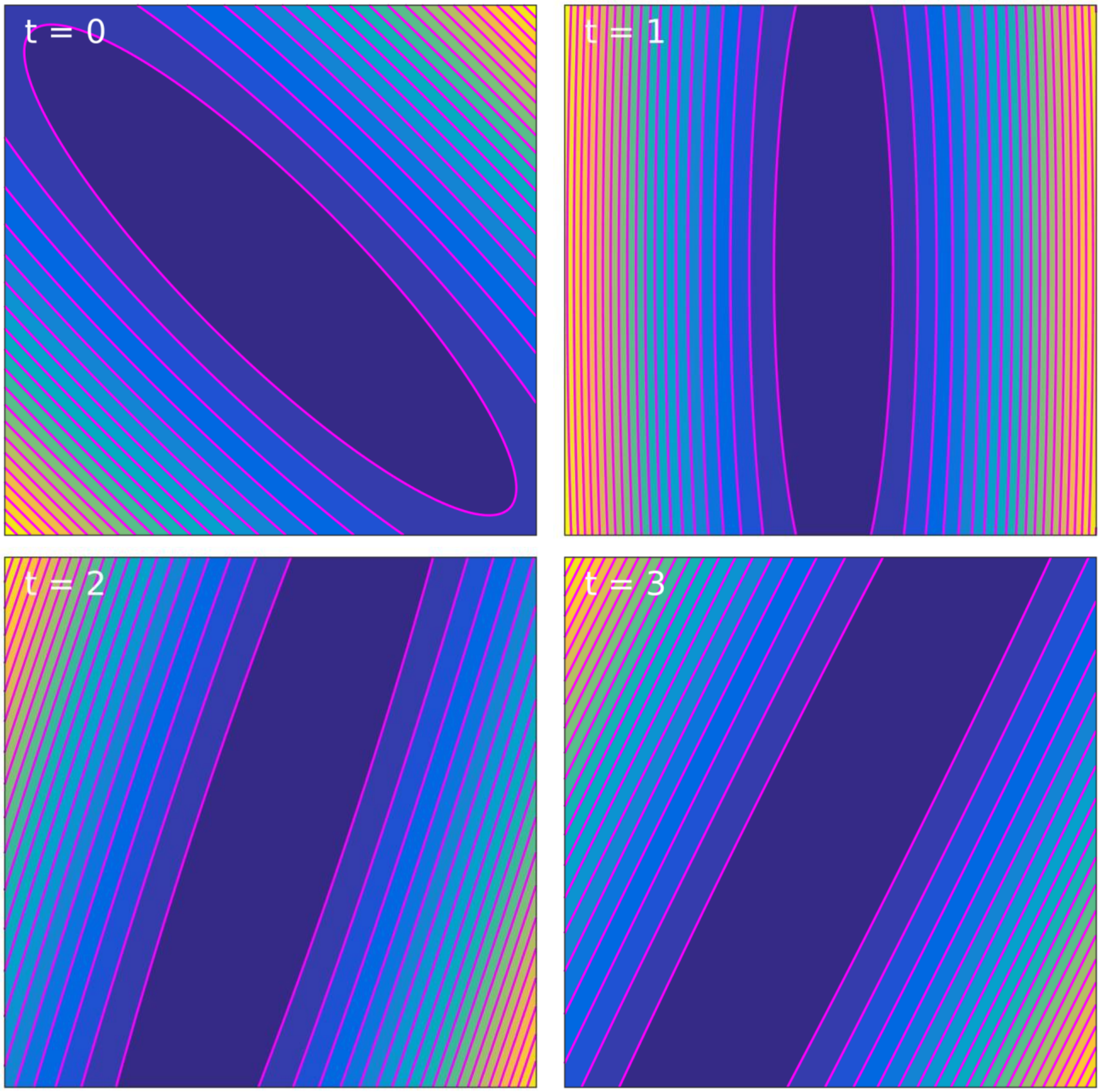

**Figure 8**: Contours of the symmetry solution of the first stream function in Eq.(40) at different time.

Notably, we can see the symmetry solutions alter the velocity fields in figure 8 and 10, suggesting symmetry solutions do have some particular importance of controlling flow patterns. The circles in the patterns of symmetry solutions indicate low velocities in the center of the circles, which are substantially different from the original patterns.

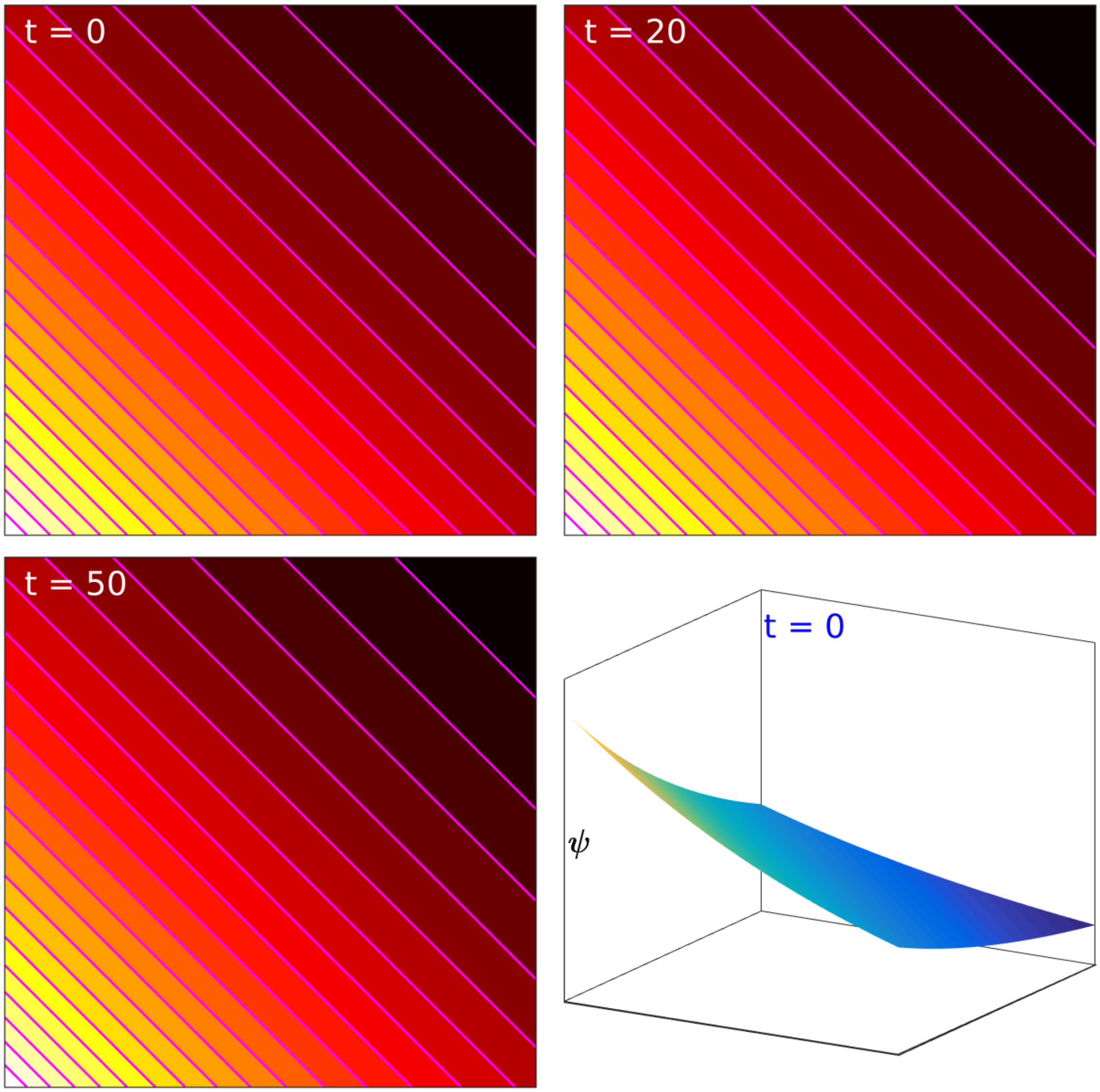

**Figure 9**: Contours of the second stream function in Eq.(40) at different time and the three-dimensional profile at t = 0.

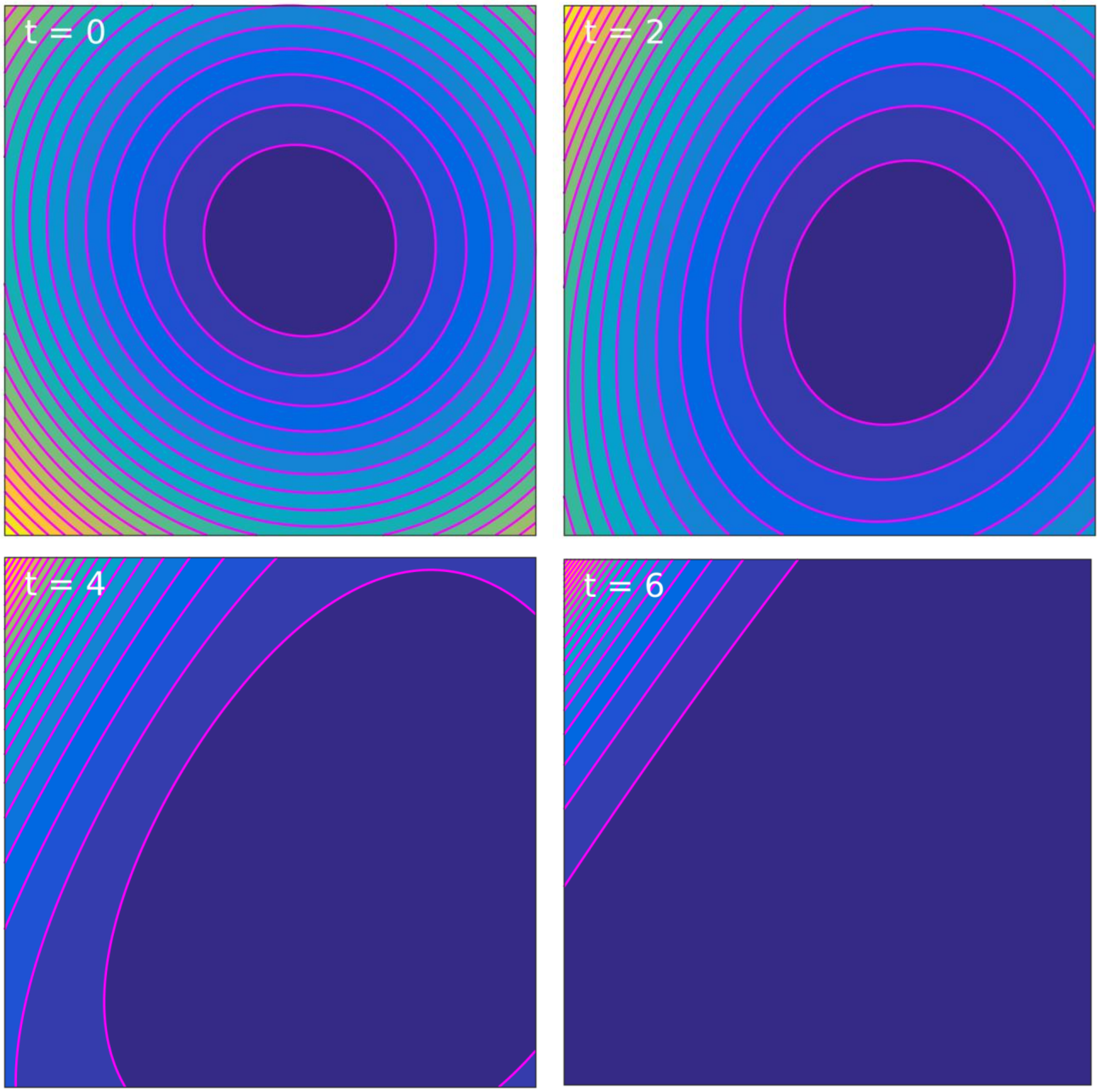

**Figure 10**: Contours of the symmetry solution of the second stream function in Eq.(40) at different time.

### *E. Symmetry solution of Lame-Oseen vortex*

As for the last case, we may see the symmetry solutions of the Lame-Oseen Vortex, which is in the form of

$$\psi(x,y,t) = \mathrm{Ei}\left(1, \frac{\mathrm{Re}\left((x-\varepsilon yt)^2 + (y+\varepsilon xt)^2\right)}{4t}\right) + \varepsilon \frac{x^2+y^2}{2} \qquad (41)$$

The streamline patterns in figure 11 are as symmetrical as that of in figure 1. But the distance between two streamlines is of not monotonic change, which is the result of superposition of two vortices with the same center but opposite directions.

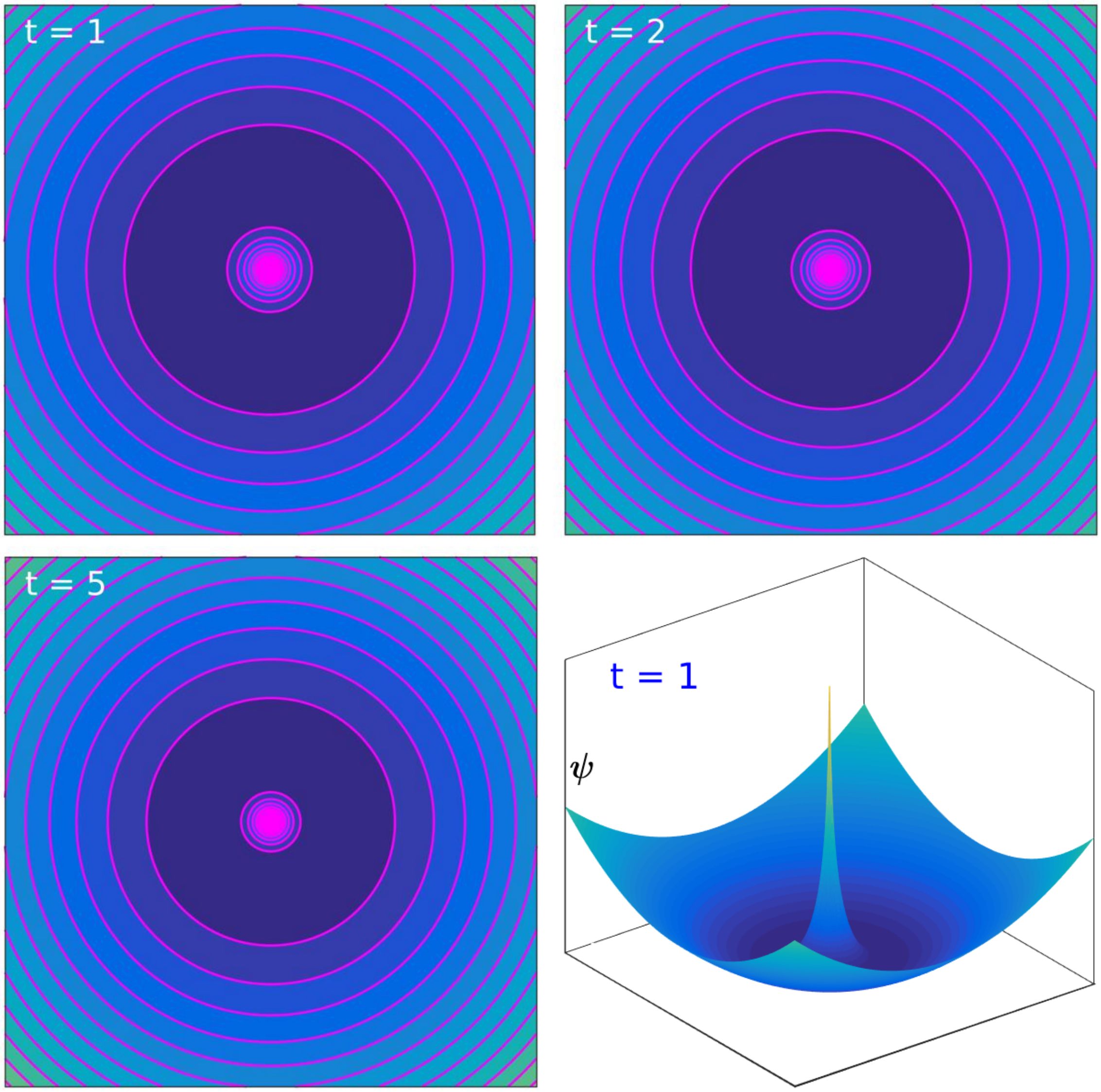

**Figure 11**: Contours of the stream function in Eq.(41) at a different time and the three-dimensional profile at t = 1.

## Conclusion

We studied the flow patterns using the traveling wave solutions of the differential equations in stream function form. Some of the traveling wave solutions have not yet been discussed in the current literature. We also showed that the traveling wave solutions are the general case of the solutions of two-dimensional incompressible potential flow that satisfy the Laplace equation. Their patterns were displayed by assuming different constant coefficients in the solution Eq.(5). The exact traveling wave solutions enabled us to characterize the flow further using their symmetry solutions transformed by Lie symmetry groups. We applied the symmetry group $G_3$, which is the group of translation, rotation and stretching to investigate flow patterns. We found that the symmetry solutions could generate complex flow patterns, e.g. the symmetry solutions of traveling solitary wave solutions created more intriguing patterns than that of the original solutions. The symmetry solution of the Lame-Oseen vortex can be seen as the outer vortex flow confines the inner vortex, such that the inner vortex is not able to diffuse out. The new exact solutions can serve as a gauge for developing numerical schemes. The symmetry solutions can also be used to control two-dimensional incompressible flows.